\documentclass[twocolumn,amsmath,amssymb,prb]{revtex4}
\usepackage{graphicx} 
\usepackage{dcolumn}
\usepackage[usenames,dvipsnames]{color}

\def \be {\begin{equation}}
\def \ee {\end{equation}}
\def \ben {\begin{eqnarray}}
\def \een {\end{eqnarray}}

\begin{document}
\author{Seogjoo Jang}
\email{seogjoo.jang@qc.cuny.edu}
\affiliation
{Department of Chemistry and Biochemistry, Queens College and the Graduate Center, City University of New York, 65-30 Kissena Boulevard, Queens, New York 11367-1597}
\author{Andr\'{e}s Montoya-Castillo}
\affiliation{Department of Chemistry, Columbia University, 3000 Broadway, New York, New York, 10027, USA}

\date{Published in the {\it Journal of Physical Chemistry B}  {\bf 119}, 7659-7665 (2015)}

\title[An \textsf{achemso} demo]{Charge Hopping Dynamics along a Disordered Chain in Quantum Environments: Comparative Study of Different Rate Kernels }




\begin{abstract}
This work presents a computational study  of charge hopping dynamics along a one dimensional chain with Gaussian site energy disorder and linearly coupled quantum bath.   Time dependent square displacements are calculated directly from numerical solutions of Pauli master equations, for five different rate kernels: exact Fermi golden rule (FGR) rate expression,  stationary phase  interpolation (SPI) approximation, semiclassical (SC) approximation, classical Marcus rate expression, and Miller-Abrahams expression.  All results demonstrate diffusive behavior in the steady state limit.  The results based on the FGR rate expression show  that the charge transport in quantum bath can be much more sensitive to the disorder than the prediction from the classical Marcus expression.  While the SPI approximation captures this general trend reasonably well, the SC approximation tends to be unreliable at both quantitative and qualitative levels, and becomes even worse than the classical Marcus expression under certain conditions.  These results offer useful guidance in the choice of approximate rate kernels for larger scale simulations, and also demonstrate significant but fragile positive effects of quantum environments on the charge hopping dynamics. 

 \end{abstract}
 \maketitle
 
 \section{Introduction}
Charge transport (CT) through disordered media plays a central role in a wide range of physical and biological processes, and has been subject to a long history of theoretical investigation and computational modeling.\cite{newton-cr91,adv-et,may,mott,coropceanu-cr107,thouless-pr13,nitzan-arpc52,newton-tca110}  While a reliable theoretical description of CT may ultimately require a detailed atomistic level calculation, such a task with satisfactory accuracy and full account of disorder is yet beyond the reach of current computational capability except for small systems.  Thus, phenomenological or semi-phenomenological approaches incorporating some microscopic information persist to be  practical methods of choice. This is true in particular for mesoscale charge transport in disordered organic molecular environments (OME) such as organic light emitting diode (OLED),\cite{friend-nature397} organic field effect transistor (OFET),\cite{drury-apl73} and organic photovoltaic (OPV) devices.\cite{deibel-rpp73} Various theoretical models\cite{bassler-pss175, friederich-jctc10,kenkre-prl62,dunlap-cp178,ruhle-jctc7,yao-jcp136} of CT focused on different physical aspects of these systems are available.  

In most disordered OMEs, charge carriers are localized (to varying extents) and their dynamics at large length scales involve incoherent transfer mechanisms.  Thus, hopping models serve as natural theoretical frameworks for the description of mesoscopic CT dynamics.    One of the most well known approaches in this regard is the Gaussian disorder model (GDM),\cite{bassler-pss175} which assumes Gaussian distributions in both site energies and the convolution of electronic couplings between charge centers.  GDM has helped exploring realistic effects of disorder, the density of active transport sites, and the applied electric field, through Monte Carlo simulation.\cite{bassler-pss175} However, the choice of hopping rate and its microscopic justification has remained ambiguous.  For example, in its original implementation, the following Miller-Abrahams (MA) form\cite{miller-pr120} was used for the hopping rate:
\be
k^{MA}_{n\rightarrow m}=k_0\exp \left (-\frac{\Delta G_{mn}}{2k_BT}-\frac{|\Delta G_{mn}|}{2k_BT}\right )\ , \label{eq:k_ma}
\ee 
where  $k_0$ decays exponentially with the distance between the two sites $n$ and $m$, and $\Delta G_{mn}=G_m-G_n$ is the Gibbs free energy of site $m$ relative to $n$. The MA form is based on the assumption that the localization of charge carrier is dominated by disorder and that the electron-phonon coupling is weak.\cite{yang-acn2,watkins-nl5,deibel-prl103} However, in soft environments such as OMEs, the coupling of charge carriers to molecular vibrations and phonon modes are not weak in general and may even dictate charge localization.  Later works\cite{parris-prl87,fishchuk-pss3,fishchuk-prb76,freire-prb72,frost-nl6,groves-jap105} accounted for this issue by employing the following Marcus rate expression:\cite{marcus-jcp24,marcus-jcp43}
\be
k^M_{n\rightarrow n}=\frac{J_{nm}^2}{\hbar^2} \left (\frac{\pi\hbar^2}{k_BT \lambda}\right)^{1/2} \exp \left (-\frac{(\Delta G_{mn}+\lambda)^2}{4\lambda k_BT}\right )\ ,  \label{eq:k_m}
\ee
where $J_{nm}$ is the electronic coupling between sites $n$ and $m$, which decays exponentially with the distance, and $\lambda$ is the reorganization energy representing the response of the environment.    

While the Marcus rate expression accounts for the effects of the  electron-phonon coupling, it relies on the assumption that all the phonon modes behave classically.  However, it is a well established fact that the quantum effects of vibrational modes make significant contributions to electron transfer (ET) reactions in molecular systems.\cite{newton-cr91,adv-et,closs-science240,jortner-jcp64,siders-jacs103,siders-jacs103-2,vanduyne-cp5,gehlen-jcp97,tang-jcp99,lang-cp244,jang-jpcb110}  Similar nuclear quantum effects can be important for CT in OME, considering  that the range of typical vibrational spectra of organic molecules is much larger than thermal energy.   Indeed, Asadi and coworkers recently showed that proper quantum mechanical treatment of vibrational modes is crucial for correct description of the current-voltage characteristics in poly(3-hexylthiophene-2,5-diyl) (P3HT) conjugated polymer systems, which are used as active organic materials in ferroelectric field effect transistors.\cite{asadi-nc4}

In general, CT processes in disordered OMEs are more complex than ET reactions in molecular systems.  The extent of charge localization is often unknown.  The details of coupling between charge carriers and their environments such as local molecular vibrational modes,  nonlocal phonon modes, and polarization response,  can be very complicated.     Nevertheless, simple models with appropriate features of OMEs can help gain deeper insights and can be used for more comprehensive computational study.  In particular, a quantitative understanding of the effects of quantum environments and their interplay with the disorder requires an extensive set of calculations, for which choosing a minimal model is helpful. Thus,  to investigate the effects of quantum environments on the overall charge hopping dynamics, we here employ a model consisting of a one dimensional chain with Gaussian site energy disorder and linearly coupled phonon bath.  

\section{Model and Computational Methods}
Assuming a spin-boson type Hamiltonian\cite{leggett-rmp59,weiss} or resorting to the linear response formulation,\cite{georgievskii-jcp110} one can calculate the charge transfer rate between two sites, $n$ and $m$, by the following Fermi golden rule (FGR) expression:
\be
k_{n\rightarrow m}^{FG}=\frac{J_{nm}^2}{\hbar^2}\int_{-\infty}^\infty dt e^{-it\Delta G_{mn}/\hbar -\left ({\mathcal K}(0)-{\mathcal K}(t)\right)} \ , \label{eq:fg_et}
\ee 
where 
\be
{\mathcal K}(t)=\int_0^\infty d\omega {\mathcal J}(\omega) \left [\coth\left (\frac{\hbar\omega}{2k_BT}\right)\cos(\omega t)-i\sin (\omega t)\right ]\ .
\ee
In the above expression, ${\mathcal J}(\omega)$ is the spectral density of the bath, all the environmental degrees of freedom, which are assumed to respond linearly to the charge transfer from the $n$th to the $m$th site.  
The use of the Marcus rate expression,\cite{marcus-jcp24,marcus-jcp43} Eq. (\ref{eq:k_m}), amounts to the assumption that all the modes constituting the spectral density can be assumed to be classical.  In this limit, the reorganization energy, defined as $\lambda=\int_0^\infty d\omega {\mathcal J}(\omega)\hbar\omega $, serves as a single parameter representing all the environmental effects on the CT.  More realistically, it is expected that the spectral density and thus $\lambda$ is not uniform across all the sites because the local environments around each charge center can be different in OMEs.  Both Eqs. (\ref{eq:k_m}) and (\ref{eq:fg_et}), and other approximate expressions considered here can be adopted for such situations.

While direct numerical evaluation of the integral in Eq. (\ref{eq:fg_et}) is possible, having a closed form approximation is useful because it can offer better physical insight and capability for larger scale calculations.  Although Jortner's rate expression\cite{jortner-jcp64} has been used widely for ET reactions, its practical application is limited to cases with few discrete quantum modes.  Another well-known expression is the so called semiclassical (SC) approximation\cite{lax-jcp20,hopfield-pnas71} given by 
\be
k^{sc}_{n\rightarrow m}=\frac{J_{nm}^2}{\hbar^2}\left(\frac{\pi\hbar^2 }{k_BT \lambda_{q,c}}\right)^{1/2}\exp\left (-\frac{(\Delta G_{mn}+\lambda )^2}{4k_BT\lambda_{q,c}}\right ) \  \label{eq:ksc} \ ,
\ee 
where $\lambda_{q,c}$ is a quantum reorganization energy defined as
\be
\lambda_{q,c}\equiv \frac{\hbar^2}{2k_BT} \int_0^\infty d\omega {\mathcal J}(\omega) \coth \left (\frac{\hbar\omega}{2k_BT} \right ) \omega^2 \ .
\ee
However, the accuracy of Eq. (\ref{eq:ksc}) is limited to a small region around the barrierless case ($\Delta G_{mn}+\lambda=0$), and its errors can be significant in other regions.\cite{siders-jacs103,siders-jacs103-2}  In fact, simple stationary phase approximations for two other points of $\Delta G_{mn}=0, \lambda$ can be derived,\cite{vanduyne-cp5,gehlen-jcp97,tang-jcp99,lang-cp244,jang-jpcb110} which involve the following two additional quantum reorganization energies:
\ben
&&\lambda_{q, s} = \frac{\hbar^2}{2k_BT} \int_0^{\infty} d\omega\ {\mathcal J}(\omega) \frac{ \omega^2}{\sinh(\hbar\omega/2k_BT)} \ ,\\
&&\lambda_{q, t} = 4k_BT \int_0^{\infty} d\omega\ {\mathcal J}(\omega) \tanh \left (\frac{\hbar\omega}{4k_BT}\right ) \ . 
\een
Most recently, a new expression combining all three stationary phase expressions through interpolation (SPI) was developed,\cite{jang-jpcb110} which can be expressed as follows:
\be
k_{n\rightarrow m}^{spi}\approx \frac{J_{nm}^2}{\hbar^2}\left(\frac{\pi\hbar^2}{k_BT\lambda_{q,c}U_{nm}}\right)^{1/2}\exp \left (-\frac{\left (\Delta G_{mn} + \lambda \right)^2}{4k_BT\lambda_{q,c}} \right ) \ , \label{eq:ksp-int}
\ee 
where 
\be
U_{nm}=\left ((1-Q_r)\left(\frac{\Delta G_{mn}}{\lambda} \right)^2+Q_r\right ) (1+\alpha e^{\gamma \Delta G_{mn}/\lambda})\ ,
\ee
 with $Q_r =\lambda_{q,s}/\lambda_{q,c}$,  and 
 \ben
&&\alpha = \exp \left ( \frac{\beta(\lambda_{q,c}\lambda_{q,t} - \lambda^2)}{2\lambda_{q,c}}\right ) - 1, \\
&&\gamma = \ln \left  [\exp \left (\frac{2\beta \lambda (\lambda_{q,c} - \lambda)}{ \lambda_{q,c}} \right ) - 1 \right ]  - \ln \left [\alpha \right ] .
 \een
Comparison\cite{jang-jpcb110} of Eq. (\ref{eq:ksp-int}) with exact numerical evaluation of Eq. (\ref{eq:fg_et}) showed reasonably good agreement for the cases where the magnitude of $\Delta G_{mn}$ is smaller than or comparable to $\lambda$, when tested for the following standard Ohmic spectral density with exponential cutoff:
	\begin{equation}\label{Eq:SpecDen}
	{\mathcal J}(\omega) = \eta \frac{e^{-\omega/\omega_c}}{\omega}\ .
	\end{equation}
Here, $\eta$ is is a measure of the coupling between the system and the bath, and $\omega_c$ is the cutoff frequency of the bath spectral density.

\begin{figure}
\makebox[0.8in]{ }(a)\makebox[2.in]{ }(b)\makebox[3in]{ }\\
\includegraphics[scale=0.25]{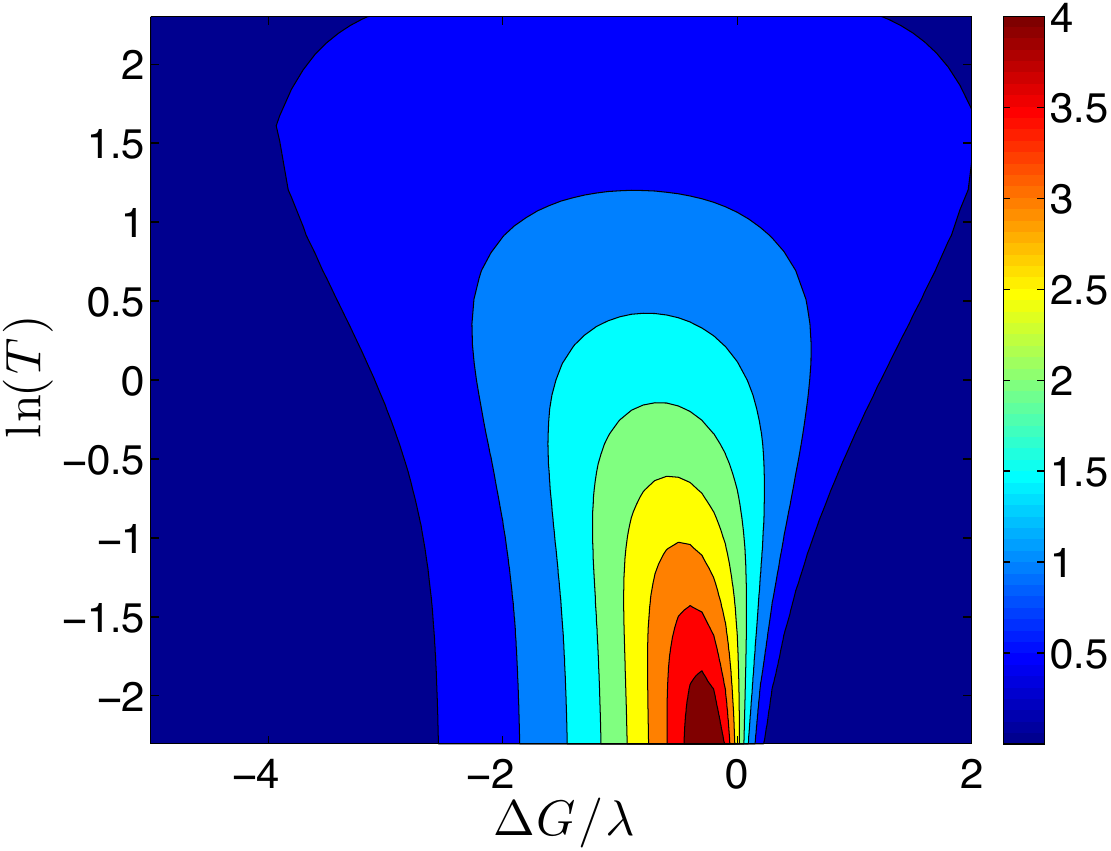}\hspace{.2in}\includegraphics[scale=0.25]{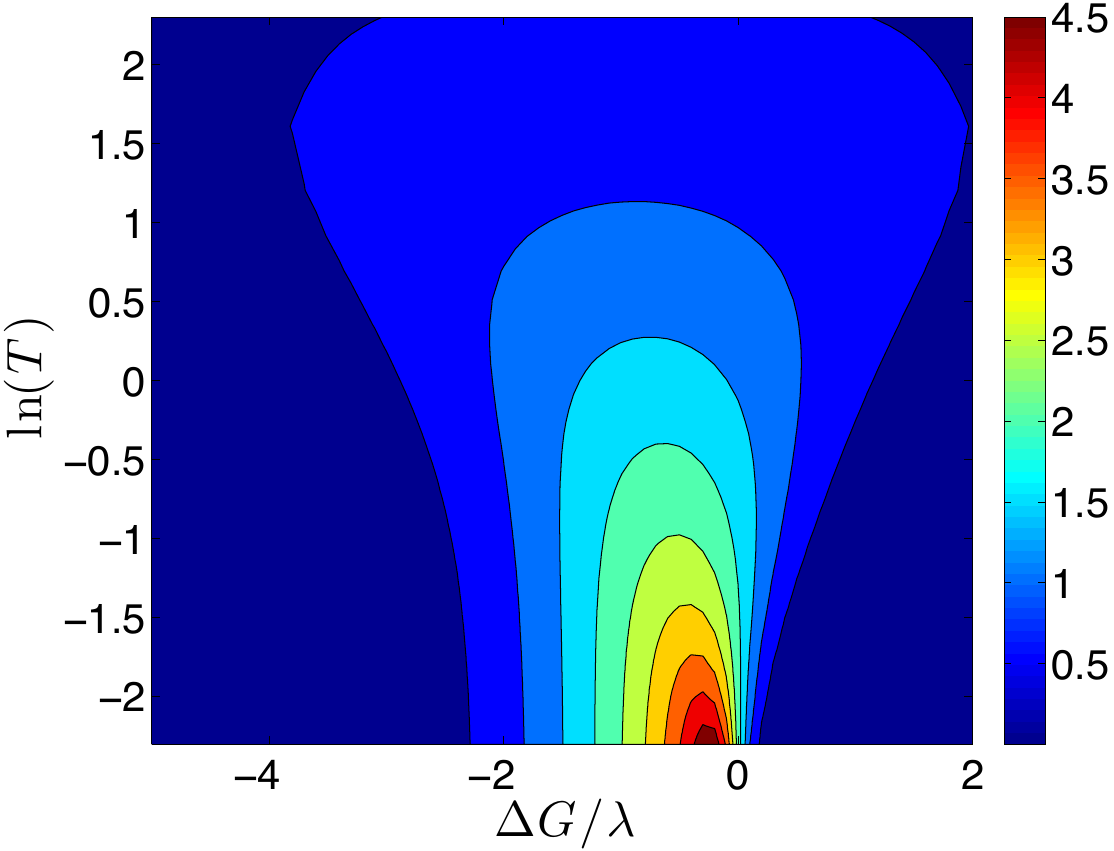}\vspace{.1in} \\
(c)\makebox[2.in]{ }(d)\makebox[2.in]{ }(e)\makebox[2in]{ }\\
\includegraphics[scale=0.25]{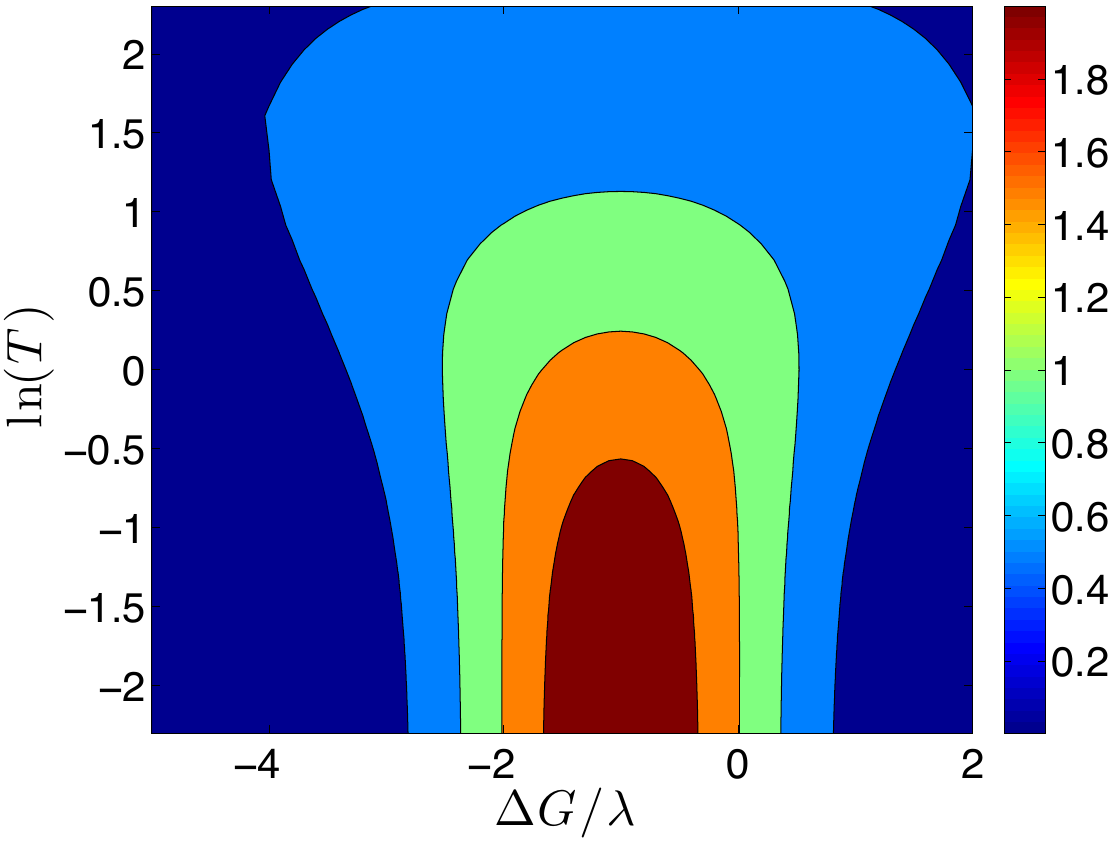}\hspace{.2in}\includegraphics[scale=0.25]{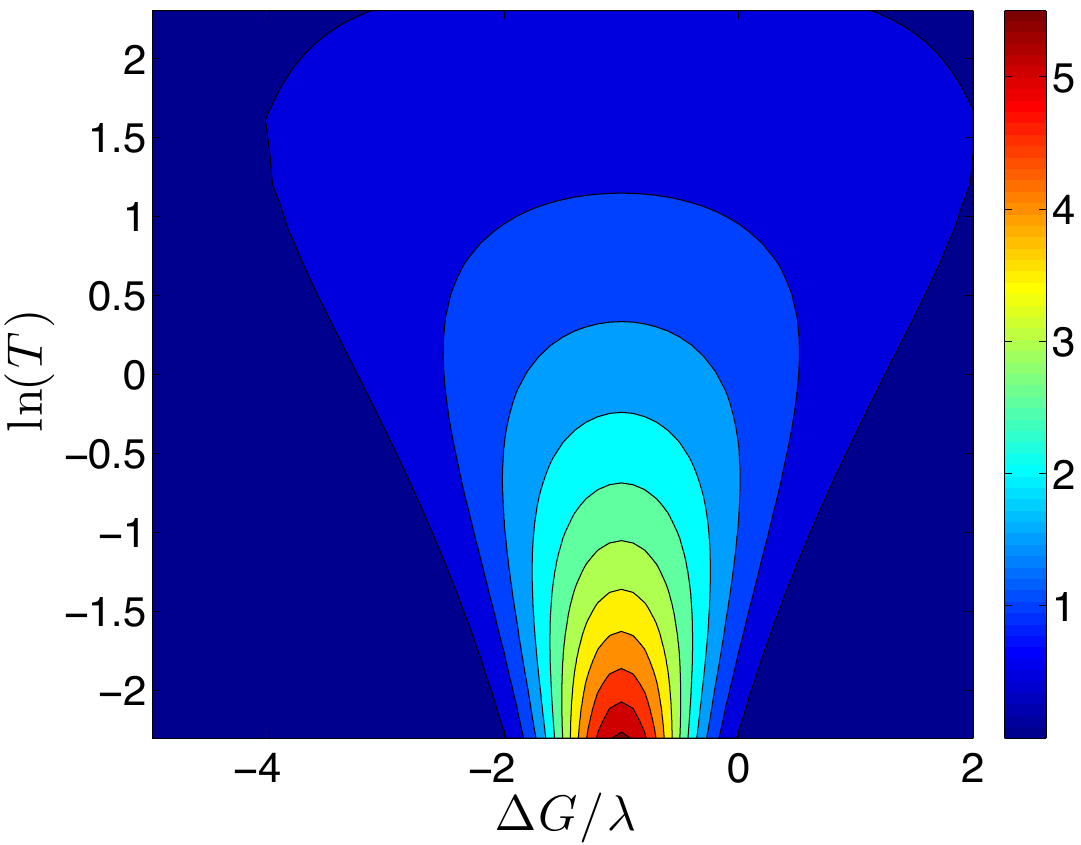}\hspace{.2in}\includegraphics[scale=0.25]{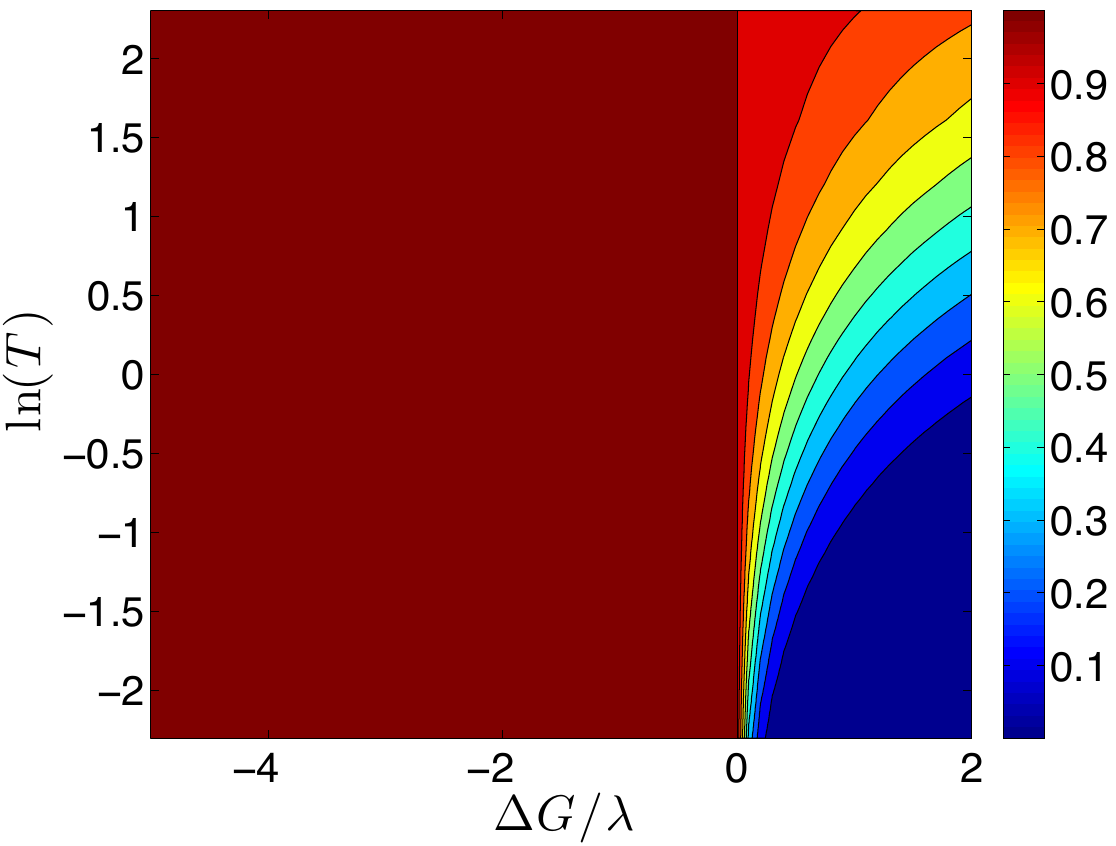}\\

\flushleft \caption{Natural logarithm of the (a) $k^{FG}$ given by Eq.~(\ref{eq:fg_et}), (b) $k^{spi}$ by Eq.~(\ref{eq:ksp-int}), (c) $k^{sc}$ by Eq.~(\ref{eq:ksc}), (d) $k^M$ given by Eq.~(\ref{eq:k_m}), and (e) $k^{MA}$ given by Eq. (\ref{eq:k_ma}) for the ranges of $\ln(T)$, where $T$ is in the unit of $k_B$, and $\Delta G/\lambda$.  It was assumed that $\eta = 1.0$ $J_{nm}=1$, and the units were chosen such that $\omega_c = 1$ and $\hbar=1$. }
\end{figure}

\ \vspace{.5in}\\

For the case where $\eta=1$ and $J_{nm}=1$, and in the units where $\hbar=1$ and $\omega_c=1$, we have evaluated the rates for a range of temperature and $\Delta G$.
Figure 1 shows two dimensional scans of the rates on the dependence of temperature  and the reaction free energy.  It is evident that the exact quantum expression for the rate predicts a strongly asymmetric pattern with respect to $\Delta G$, particularly at low temperatures, which is most accurately captured by the SPI approximation, $k^{spi}$.  On the other hand, the Marcus expression $k^M$ and the SC approximation $k^{sc}$ are accurate only at high temperatures and do not reproduce correct qualitative behavior in the low temperature limit.  As a reference, we have also calculated the MA rate expression Eq. (\ref{eq:k_ma}) assuming $k_0=1$, which fails to describe the temperature dependence in particular for the exothermic region.

We then conducted simulation of CT for a model $N$-site 1D system characterized by energetic disorder in the limit of zero applied electric field and with only nearest neighbor hopping transfers.  Thus, the site free energy $G_n$ for each $n\neq 0$, $N$ was drawn from the following Gaussian distribution:
\begin{equation}
	g_{\sigma}(G) = \frac{1}{\sqrt{2\pi \sigma^2}} \exp \left (-\frac{ G^2}{2\sigma^2}\right )\ , \label{eq:gaussian}
	\end{equation}
where $\sigma$ is the standard deviation quantifying the disorder of free energy at each site. We also assumed that $G_0=0$ and $G_N=-1$.   In general, 
the time evolution of the population of the charge carrier  at the $m^{th}$ site is described by
	\begin{equation}\label{Eq:MasterEq}
	\frac{dP_m(t)}{dt} = \sum_{n=1, n\neq m}^N  \left \{ k_{n \rightarrow m} S(P_m)P_n(t)- k_{m \rightarrow n}S(P_{n}) P_{m}(t) \right \}, 
	\end{equation}
where 	
$S(P)$ represents the self-correlation at each site.  Assuming the dilute charge carrier limit, we here ignore the effect of correlation and use $S(P) = 1$.  Thus, Eq.~(\ref{Eq:MasterEq}) takes the form of the Pauli master equation (PME), which can be solved easily.  Under the boundary conditions of $P_0 = P_{N+1} = 0$ and starting from the initial condition of $P_n(t = 0) = \delta_{n1}$, we solved Eq. (\ref{Eq:MasterEq}) for $N=100$.  The fourth-order Runge Kutta algorithm\cite{lambert} was used with a time step of $dt = 0.1$.  Although it is also possible to formally solve Eq. (\ref{Eq:MasterEq}) when written as a matrix equation, implementation of the solution requires diagonalization of the rate matrix.  This procedure becomes difficult for large system sizes and higher dimensions.  Instead, the Runge-Kutta algorithm can be easily scaled to larger systems with ease and yields the same result when implemented with a sufficiently small time step.   For all the calculations, we have assumed $\eta=1$ and $J_{nm}=1$, and the units of $\hbar=1$ and $\omega_c=1$.  In addition, the inter-site distance was assumed to be the unit distance.  

For each realization of a random wire, {\it i.e.}, a sample of site energies drawn from the Gaussian distribution of Eq. (\ref{eq:gaussian}), we solved the PME as described above and calculated the following time dependent mean square displacement: 
\be
R^2(t) = \sum_{n=1}^N P_n(t) n^2  \ .\label{eq:delta R2}
\ee
This procedure was repeated over an ensemble of 2,000 realizations of the Gaussian distributions of site energies, over which Eq. (\ref{eq:delta R2}) was averaged to calculate the average mean square displacement $\langle R^2(t)\rangle$.   These calculations were conducted for each of the rate kernels, Eqs. (\ref{eq:k_ma})-(\ref{eq:fg_et}), (\ref{eq:ksc}), and (\ref{eq:ksp-int}), at a few different values of $k_BT$ within the range of $0.1-10$. When calculated up to $t=100$, all results exhibited steady diffusive behavior.
\begin{figure}
\includegraphics[width=3 in]{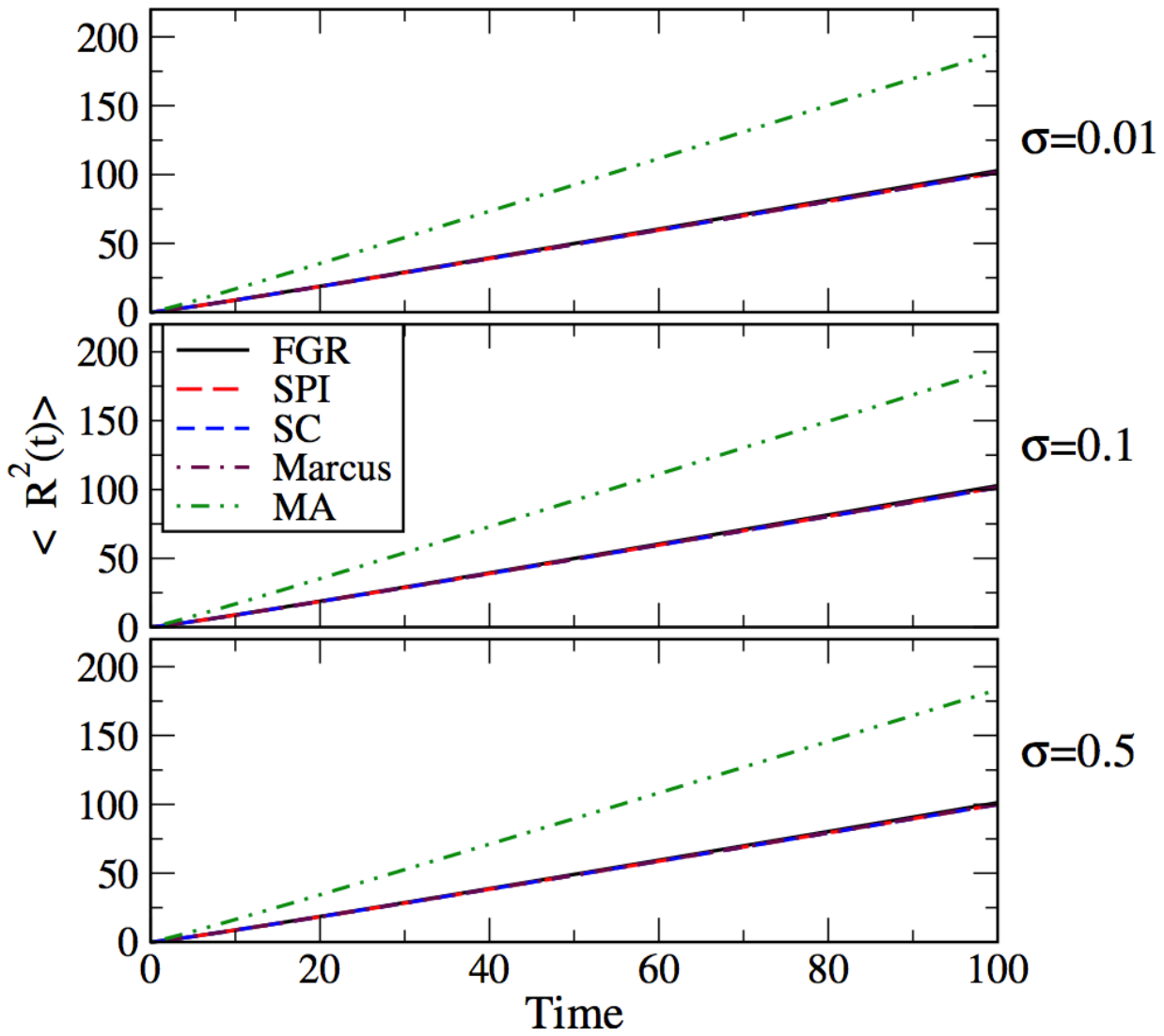}
\caption{Values of $\langle R^2(t) \rangle $ vs. time based on different rate kernels, Eqs. (\ref{eq:fg_et}) (FGR), (\ref{eq:ksp-int}) (SPI), (\ref{eq:ksc}) (SC), (\ref{eq:k_m}) (Marcus), and (\ref{eq:k_ma}) (MA) for $k_BT=10$.  Units are chosen such that $\hbar=1$, $\omega_c=1$, and the intersite distance is equal to one.       }
\end{figure}

\section{Results and Discussion}
Figure 2 shows the results for the high temperature limit, $k_BT=10$.  As expected, all results except for those of MA agree regardless of the value of $\sigma$.  This confirms that the bath can be approximated as classical at this temperature.  The MA expression is shown to systematically overestimate this result by more than a factor of two.    While this discrepancy may be adjusted by correcting the value of the prefactor $k_0$ at a particular temperature, the overall temperature dependence cannot be addressed this way as will become clear below. 

Figure 3 shows  the results at $k_BT=1$,  where the bath is close to being classical but has some quantum character.  Comparison of the FGR results with the classical Marcus results shows that the quantum bath  enhances the transport, but its positive effect diminishes as the disorder increases.   Upon close examination of the plot of the rate versus reaction free energy, we find that this is because the FGR rate is appreciably higher than the Marcus rate only for $|\Delta G|/\lambda \leq 1$.  The SPI results underestimate the exact FGR results but remain close.  The SC results also remain close to the FGR results at this temperature, but the trends are not consistent.   The MA results are significantly lower than the FGR results at this temperature, which is opposite to the trend seen in Fig. 2.  

\begin{figure}
\includegraphics[width=3 in]{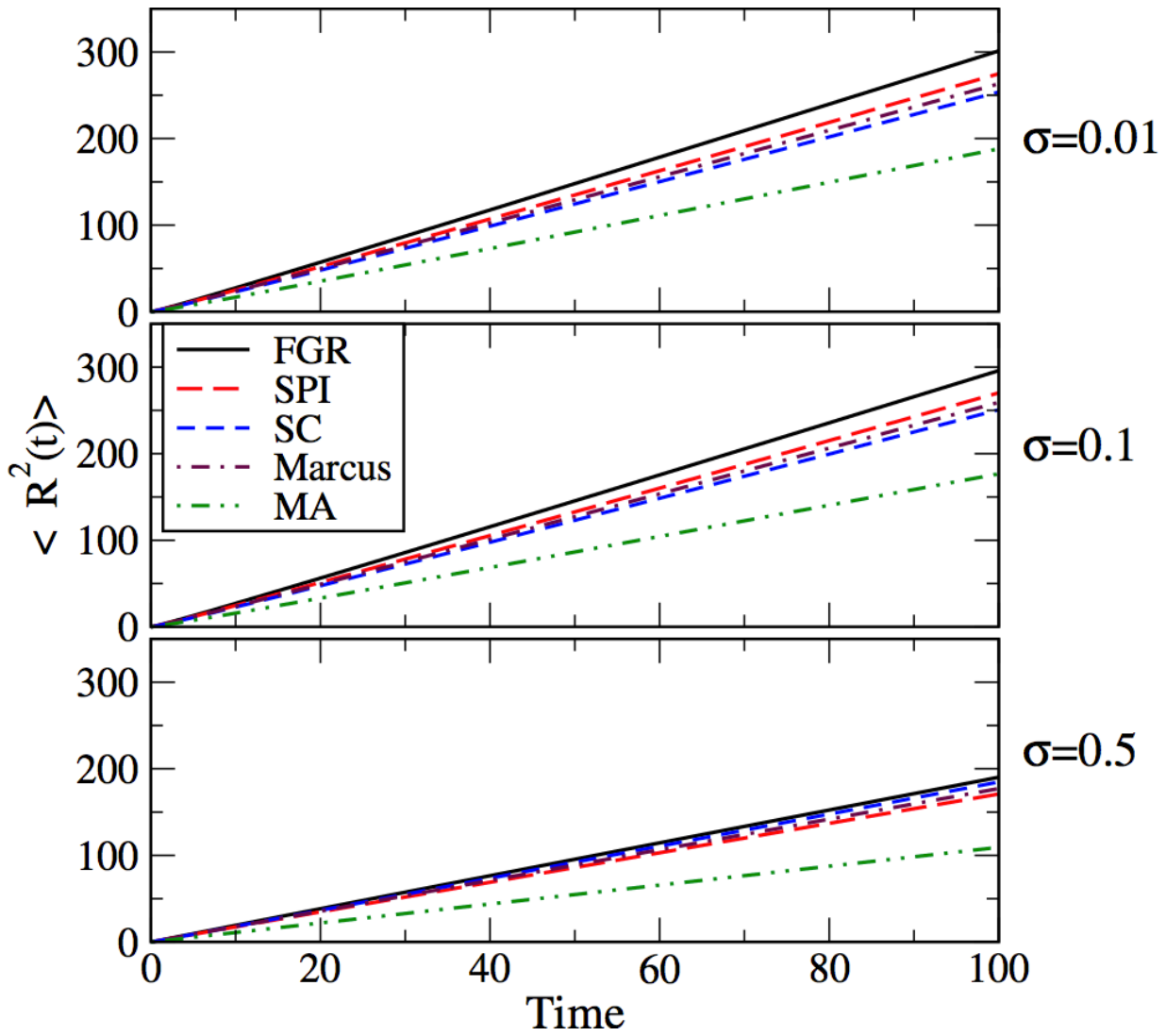}
\caption{Values of $\langle R^2(t) \rangle $ vs. time based on different rate kernels, Eqs. (\ref{eq:fg_et}) (FGR), (\ref{eq:ksp-int}) (SPI), (\ref{eq:ksc}) (SC), (\ref{eq:k_m}) (Marcus), and (\ref{eq:k_ma}) (MA) for $k_BT=1$. Units are chosen such that $\hbar=1$, $\omega_c=1$, and the intersite distance is equal to one. }
\end{figure}

Figure 4 shows the results for the low temperature limit $k_BT=0.1$, where the bath becomes fully quantum mechanical.  Comparison of the FGR and Marcus results show that the quantum effects of the bath can accelerate the transport substantially as can be seen for $\sigma=0.01$, where the enhancement factor is more than four.  However, such positive effects of quantum bath diminish quickly as the disorder increases.  At the value of $\sigma=0.5$, the discrepancy between the quantum FGR and classical Marcus results has become very small.  This demonstrates the fragile nature of quantum effects on one hand and also shows that the use of the classical Marcus expression in this regime of disorder is well justified.  Even the MA result agrees well with the FGR result, which suggests that the hopping dynamics is dominated by the disorder.   While the SPI approximation remains close to the exact FGR result, the SC approximation does not.  In fact, the error of the latter is the worst among all approximations.  This can be understood from the fact that the SC approximation substantially overestimates the quantum effect in the normal or deep inverted regime.\cite{jang-jpcb110}

\begin{figure}
\includegraphics[width=3 in]{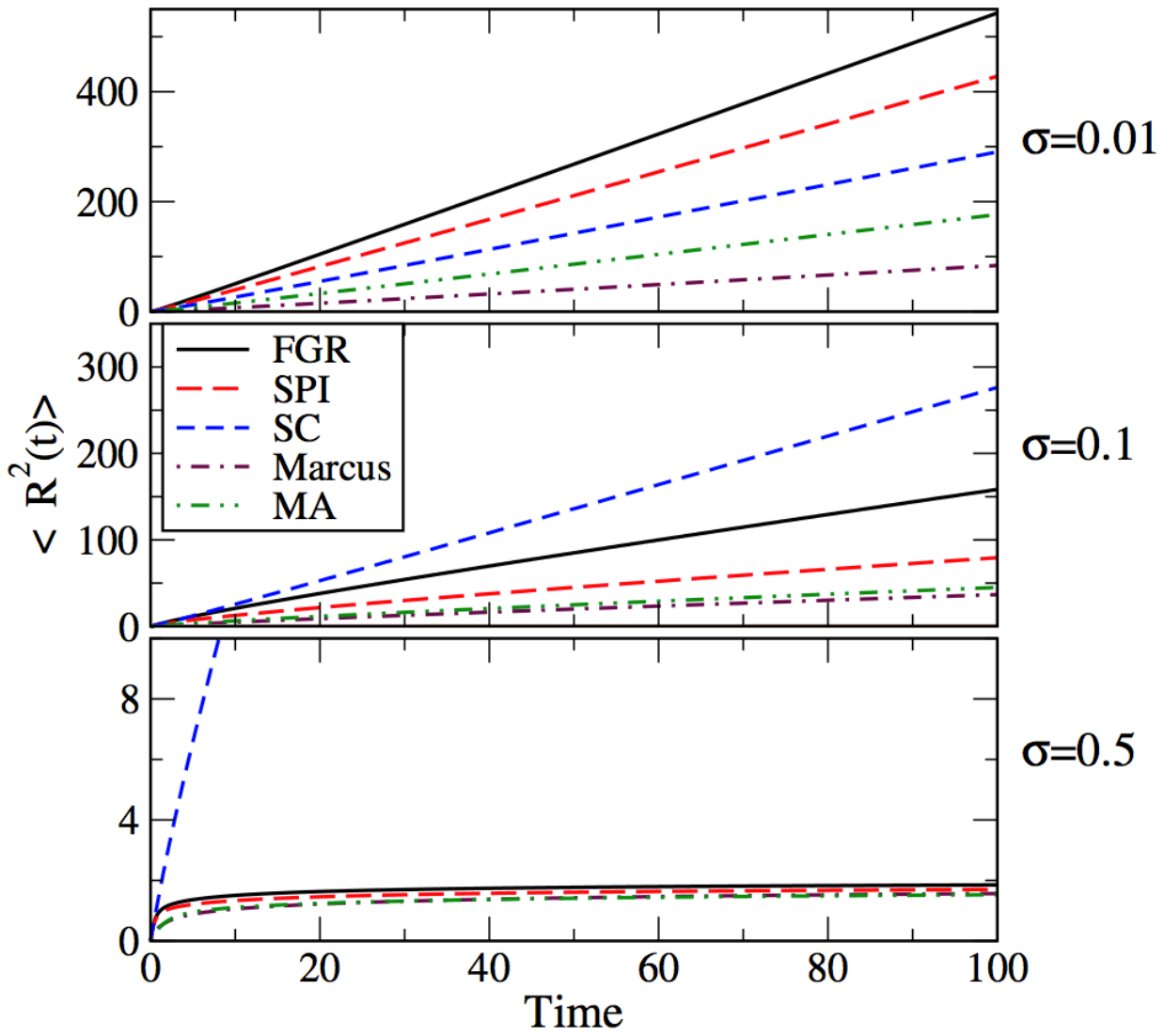}
\caption{Values of $\langle R^2(t) \rangle $ vs. time based on different rate kernels, Eqs. (\ref{eq:fg_et}) (FGR), (\ref{eq:ksp-int}) (SPI), (\ref{eq:ksc}) (SC), (\ref{eq:k_m}) (Marcus), and (\ref{eq:k_ma}) (MA) for $k_BT=0.1$. Units are chosen such that $\hbar=1$, $\omega_c=1$, and the intersite distance is equal to one. }
\end{figure}

As a side note, we want to clarify that the set of parameters used in Fig. 4, where substantial quantum effects of the bath can be seen, are not far from real situations of OMEs.   In a recent simulation that models charge transport in bulk heterojunction (BHJ) morphology,\cite{groves-jap105} it was assumed that $\lambda \sim 0.25\ {\rm eV}$ and $\sigma=0.1\ {\rm eV}$.  At room temperature ($k_BT\approx 0.025\ {\rm eV}$) and assuming  $\hbar\omega_c=0.18\ {\rm eV}$, which is the approximate vibrational energy of ${\rm C=C}$ double bond, such parameters correspond to $\eta=1.4$, $\Delta G/(\hbar\omega_c)=0.55$, and $k_BT/(\hbar\omega_c)=0.14$.   The temperature corresponding to this in our units and the choice of $\eta=1$, corresponds to about $k_BT=0.2$, for which the results are similar to  those in Fig. 4.   

In order to relate out results to actual transport properties of charge carrier, for all the results shown in Figs. 2-4, we have estimated relevant diffusion constants as follows:
\be
D= \lim_{t\rightarrow \infty} \frac{d\langle R^2(t)\rangle}{2dt} \approx \left . \frac{d \langle R^2(t) \rangle}{2dt}\right |_{t=100} \ .
\ee 
We have also conducted similar calculations for other values of $k_BT=0.2$, $0.5$, $1$, and $2$, and calculated the diffusion constants.   Figure 5 shows the resulting values with the variation of temperature for three different values of $\sigma$, which provide the overall trends more clearly. 

\begin{figure}
\includegraphics[width=3in]{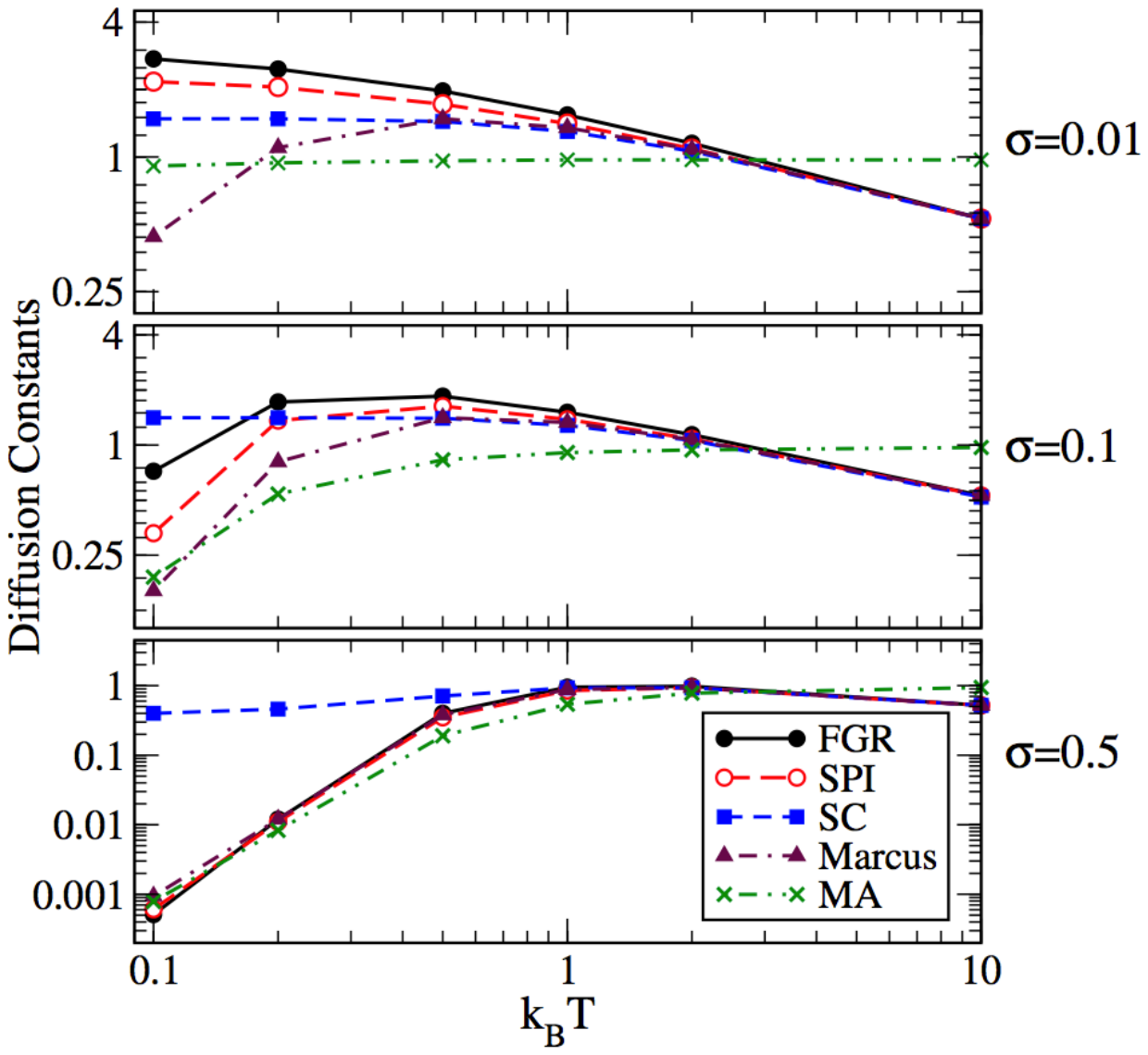}
\caption{ Temperature dependences of diffusion constants extracted numerically from the asymptotic slopes calculated for $k_BT=0.1$, $0.2$, $0.5$, $1$, $2$, and $10$.  The lines are drawn as guide.  Both axes are shown in logarithmic scales.}
\end{figure}

In the limit of very weak disorder ($\sigma=0.01$),  lowering of temperature (within the range being considered) results in enhancement of the diffusion constant.  This trend is reproduced best by the results of SPI.   For $k_BT\leq 0.5$, the SC results become insensitive to temperature and the classical Marcus results show an opposite trend.  The MA results are almost insensitive to temperature throughout the region considered.   

For moderately weak  disorder ($\sigma=0.1$),  the qualitative behavior of FGR results become closer to the classical Marcus results, although the positive effects of quantum bath are still significant in the low temperature limit.  The SPI results capture this trend better than any other approximations, but become less accurate in the low temperature limit.   The SC results exhibit a qualitative trend different from that of the FGR results in the low temperature limit.  The MA results do not show correct temperature dependence either. 

For moderately strong disorder ($\sigma=0.5$), all the FGR, SPI, and the classical Marcus results are very close.  Even the MA approximation approaches the FGR result in overall temperature dependence, although significant discrepancy can still be seen at smaller scale.   This indicates that the hopping dynamics is mostly dominated by the disorder, as can be seen from the relatively good performance of the MA approximation.  On the other hand, the SC approximation substantially overestimates the quantum effects and becomes unreliable in the low temperature limit.  This is consistent with the previous assessments\cite{siders-jacs103,bader-jcp93} that the performance of SC can even be worse than that of the classical Marcus expression.

\section{Conclusion}

In summary, we have tested the effects of different rate kernels on the  CT in a one dimensional chain with Gaussian disordered site energies and  linearly coupled quantum bath, by utilizing numerically exact solution of the Pauli master equation for the time dependent population.  All calculation results demonstrated  diffusive behavior of charge carriers in the long time limit.  We confirmed that the FGR results and all of their approximations converge in the high temperature limit (Fig. 2), where the bath can be assumed to be classical. 
At lower temperatures where the thermal energy is comparable to or smaller than the reorganization energy of the quantum bath, the classical Marcus expression results in underestimation of the the exact hopping dynamics based on FGR, with varying degree depending on the disorder.  This indicates that quantum bath has positive effect on the hopping dynamics of charge carriers in general.  However, as can be seen in the bottom panel of Fig. 5, such effects diminish fairly quickly as the disorder increases.  This demonstrates the fragile nature of positive quantum effects and the importance of controlling disorder for their utilization.    

Based on the consideration of all the model studies conducted here, we suggest the following order of performance among the different approximations: SPI $>$ Marcus $>$ MA, SC.   We ranked MA and SC expressions equally because the deficiencies of them are not comparable. The MA expression does not provide proper temperature dependence in the weak or moderate disorder limit, and the SC approximation becomes unreliable in the low temperature limit. On the other hand, the SPI approximation has been shown to serve as a reasonable approximation for reproducing correct qualitative dependence on temperature and disorder.  While further theoretical effort is needed to make it more accurate quantitatively, numerical tests shown here establish the SPI as a reasonable approximation for investigating the quantum effects for more realistic and large scale models.   These include models with distance dependent non-nearest neighbor hopping, driven hopping dynamics in the presence of an applied electric field, models with disorder in the spectral density and electronic couplings as well as site energies, and  hopping dynamics in two and three dimensions.  Quantitative understanding of the effects of quantum environments and their interplay with the disorder in these models can aid in the design of novel organic materials utilizing positive quantum effects for efficient CT in a robust manner.

\acknowledgements
SJ is indebted to numerous illuminating discussions with Marshall Newton and John Miller through out his career, and thanks their generosity in sharing their insights.  This work was supported by the Office of Basic Energy Sciences, Department of Energy (Grant No. DE-SC0001393),  the National Science Foundation  (Grant Nos. CHE-0846899 and CHE-1362926), and the Camille Dreyfus Teacher Scholar Award.   Part of this research was conducted during sabbatical stay of SJ at Columbia University, which was supported by the Center for Re-Defining Photovoltaic Efficiency Through Molecular Scale Control, an Energy Frontier Research Center funded by the US Department of Energy, Office of Science, Basic Energy Sciences under Award no. DE-SC0001085 and also by a matching grant from the Empire State Development's Division of Science, Technology and Innovation (NYSTAR).   We also thank David R. Reichman for a warm hospitality during SJ's stay and for the financial support of AMC.

\providecommand{\latin}[1]{#1}
\providecommand*\mcitethebibliography{\thebibliography}
\csname @ifundefined\endcsname{endmcitethebibliography}
  {\let\endmcitethebibliography\endthebibliography}{}

\end{document}